\documentclass[useAMS,usenatbib]{mn2e}
\usepackage{amssymb,amsmath}
\usepackage{amsfonts}
\usepackage{epsfig}
\usepackage{graphicx}
\usepackage{color}

\numberwithin{equation}{section}

\newcommand{\bea}{\begin{eqnarray*}}
\newcommand{\eea}{\end{eqnarray*}}
\newcommand{\ba}{\begin{array}}
\newcommand{\ea}{\end{array}}
\newcommand{\p}{\partial}

\newcommand{\ee}{\end{equation}}
\newcommand{\be}{\begin{equation}}

\newcommand{\bA}{{\mbox{\boldmath $A$}}}
\newcommand{\bB}{{\mbox{\boldmath $B$}}}

\newcommand{\bell}{{\mbox{\boldmath $\ell$}}}

\def\sl{\sum\limits}

\newcommand{\case}{\textstyle\frac}

\title[On the radial orbit instability]{On the nature of the radial orbit instability
in spherically symmetric collisionless stellar systems}
\author[E. V. Polyachenko et al.]
   {E.~V.~Polyachenko,$^1$\thanks{E-mail: epolyach@inasan.ru}
       I. G. Shukhman$^2$\thanks{E-mail: shukhman@iszf.irk.ru}\\
      $^1$Institute of Astronomy, Russian Academy of Sciences, 48 Pyatnitskya St., Moscow 119017, Russia\\
      $^2$Institute of Solar-Terrestrial Physics, Russian Academy of Sciences,
      Siberian Branch, P.O. Box 291, Irkutsk 664033, Russia}
\date{Accepted \qquad
     Received }

\pagerange{\pageref{firstpage}--\pageref{lastpage}} \pubyear{2015}

\begin{document}
\maketitle

\label{firstpage}

\begin{abstract}

We consider a two-parametric family of radially anisotropic
models with non-singular density distribution in the centre. If
highly eccentric orbits are locked near the centre, the
characteristic growth rate of the instability is much less than the
Jeans and dynamic frequencies of the stars (slow modes). The instability occurs
only for even spherical harmonics and the perturbations are purely
growing (aperiodic). On the contrary, if all orbits nearly reach the
outer radius of the sphere, both even and odd harmonics are
unstable. Unstable odd modes oscillate having characteristic
frequencies of the order of the dynamical frequencies (fast modes). Unstable
even harmonics contain a single aperiodic mode and several
oscillatory modes, the aperiodic mode being the most unstable.

The question of the nature of the radial orbit instability
(ROI) is revisited. Two main interpretations of ROI were suggested in
the literature. The first one refers to the classical Jeans instability
associated with the lack of velocity dispersion of stars in the
transverse direction. The second one refers to Lynden-Bell's orbital
approach to bar formation in disc galaxies, which implies slowness
and bi-symmetry of the perturbation. Oscillatory modes, odd spherical harmonics modes, and
non-slow modes found in one of the models show that the orbital interpretation is not the only possible.


\end{abstract}


\section{Introduction}

Spherical systems with predominance of eccentric orbits are subject
to the so-called radial orbit instability (ROI) that leads to
formation of non-spherical structures. These structures are
naturally associated with the triaxial bulges and bars, observed in
a variety of self-gravitating systems, where the eccentric orbits
could occur as a result of the radial collapse in early stages of
formation. Numerical simulations of collapsing systems have been
carried out in Aguilar and Merritt (1990), Roy and Perez (2004),
Trenti and Bertin (2006). In addition to these non-equilibrium systems, it is of interest to
study the stability of \textit{equilibrium} models that are used in
modelling galaxies, globular and open clusters. The stability
conditions may impose substantial restrictions on the allowed
parameters of models.

By analogy with the well-known Ostriker\,--\,Peebles stability
criterion for disk systems, Polyachenko and Shukhman (1981) proposed
a global stability criterion for spherical systems, {$ \zeta_{\textrm{crit}} =
2T_r / T_{\perp}> 1.7 \pm 0.25 $}, where $T_r$ and $T_{\perp}$ are
total kinetic energy of radial and transverse motion. Subsequently
it was found that the specified range for
$\zeta_{\textrm{crit}}$ is not rigorously found. In particular,
a critical value of $\zeta$ for generalized polytropic models was
found close to 1.4 (Fridman and Polyachenko, 1984, Barnes et al.
1986), or even 1 (Palmer and Papaloizou, 1987).\footnote{Note that
the result by Palmer and Papaloizou was questioned by Polyachenko et
al. (2011).} On the other hand, Osipkov\,--\,Merritt models give
examples of systems that preserve initial spherical shape with $
\zeta_{\textrm{crit}}$ as much as $2.5 $ (Meza and Zamorano 1997).
The most stable radially anisotropic configuration
($\zeta_{\textrm{crit}} \approx 2.9 $) were obtained by Trenti and
Bertin (2006) in the numerical simulation of collisionless collapse.
In the latest models, the stabilizing effect was due to the nearly
isotropic core, while large anisotropy was achieved due to a
strongly anisotropic shell.

Several mechanisms were proposed to explain ROI, among which we
mention two. The first mechanism treats this instability as the
Jeans instability of anisotropic medium, in which the velocity
dispersion in the transverse direction cannot resist gravitational
attraction (Polyachenko and Shukhman, 1972, 1981; Antonov, 1973, Barnes et
al. 1986).  The other mechanism supported by Merritt (1987, 1999),
Saha (1991), Weinberg (1991), Palmer (1994), and others is claimed to be similar
to bar formation in rotating discs described by Lynden-Bell
(1979): this is a {\it tendency} of some orbits to line up in the
direction of the bar that leads to increase of the density and
potential perturbation.

A detailed exposition of the latter approach is given by Palmer (1994), who describes allowed mathematical simplifications of a general matrix equation for eigenoscillations to study instability under question. It was assumed that:
\begin{itemize}
	\item The modes are even (even spherical harmonics $ l $),
since the stellar orbits are symmetric with respect to the centre of the system.
	\item The modes are `slow', i.e. the modes eigenfrequency
$ \omega $ must be much smaller than the characteristic
radial frequency of stars, $ | \omega | \ll \Omega_{1} $. Under this condition, periods of stars will be much shorter than characteristic
time of the instability, and the orbits can be regarded as separate objects.
\end{itemize}
These two features certainly narrow the range of possible unstable modes compared to the more general former approach, in which the frequencies $\omega$ can be of the order of the radial frequency, $ | \omega | \sim \Omega_{1} $. For convenience, we denote the two mechanisms as `Jeans' and `Lynden-Bell', but the primary difference is fastness and slowness of unstable modes. Besides the mentioned above restrictions, the `Lynden-Bell' mechanism assumes that a radial part $ \chi (r) $ of the eigenfunction of the perturbed potential $\delta \Phi (r, \theta, \varphi) = \chi (r) \, Y_l ^ m(\theta, \varphi) $ is nodeless.

In what follows we analyse the spectra of radially-anisotropic DFs of the form $F(E,L)$ to answer the question whether only slow modes, or both types of modes are possible. We imply that all unstable non-radial modes are due to ROI, provided they are stabilized by decreasing the radial anisotropy. For the analysis we employ a two-parametric family of radially anisotropic models without central singularity (Polyachenko et al.
2013):
\begin{align}
F (E, L) &= \frac{N (q, L_T)}{4\pi^3 L_T ^ 2} \, H (L_T-L) F_0 (E) \ ,  \nonumber\\
F_0 (E)  &= 2(1+q) (-2E) ^{q} \ ,
\label{apm}
\end{align}
where $ H (x) $ is the Heaviside step function; parameters $ q \geq
-1 $, $ L_T \geq 0 $; $ E $ is the energy and $ L $ is the absolute
value of the angular momentum of individual stars,
$$
E = \frac{1}{2} (v_r ^ 2 + v_{\perp} ^ 2) + \Phi_0 (r) \ , \quad L = r v_{\perp}\ .
$$
In the distribution function (DF) (\ref{apm}), the gravitational potential $ \Phi_0 (r) $ is set to zero on the sphere boundary; the boundary radius $R$, and the total mass $M$ are unity; $ N (q, L_T)$ is the normalization constant.

In contrast to the well-known Osipkov\,--\,Merritt models
(Osipkov, 1979; Merritt, 1985), and generalized polytropic models (Cumm, 1952)
\begin{align}
 F_{\rm GP} (E, L) = C (s, q) \, L ^{- s} (-2E) ^ q \ ,
\label{opm}
\end{align}
where $-1 \le q \le 7/2$, $- \infty <s <2$, our family of models
(\ref{apm}) have two advantages needed for correct analysis of the
ROI: (i) the central density and the potential are finite, and (ii)
for a wide range of parameter $q$, variation of parameter $ L_T$
transforms the system from isotropic to purely radial. A
technical advantage is that the specific form of the $ L
$-dependence of the DF significantly simplifies a cumbersome
numerical procedure of finding the eigenmodes.

Below we  investigate the stability of two series of DF (\ref{apm})
with fixed values of the parameter $ q = 0$ and $ q = -1 $.
A principal difference between the two is in the energy
distribution of stars: the former one holds equipartition, while the
latter is mono-energetic, since  $\lim\limits_{q\to -1^+} F_0(E) = \delta(E)$ (see, e.g., Gelfand and
Shilov, 1964). For highly radially anisotropic systems, models of $ q\!=\!0 $ series have
orbits with small apocentric distances (short needles) confined in
the centre, where the characteristic orbital frequency $ \Omega_{\rm
dyn} $ is very large. In contrast, in mono-energetic models length
of all  highly eccentric orbits is nearly equal, and all stars
can reach the outer radius of the system. We find that this
feature results in a completely different character of the
instability.

The plan of the paper is as follows. In Section 2 we provide a
general matrix equation to determine the eigenfrequencies, and its
special form for the series under consideration. In Section 3 we
present results of calculations of eigenmodes. Finally, in section 4 we
summarize the results and discuss the physical mechanisms of radial
orbit instability.

\section{Matrix equations}

In this paper, we address the problem of stability by finding
the eigenfunctions $ \chi (r) $ and the eigenfrequencies $ \omega $
of collisionless systems using matrix method. The method was first
proposed by Kalnajs (1977) for disk systems. For spherical systems
we are interested in a similar matrix equation, which was first obtained by
Polyachenko and Shukhman (1981). Details of the derivation has been
repeatedly given in the literature (see, e.g., Polyachenko and
Shukhman, 1981; Weinberg, 1991; Bertin et al., 1994; Saha, 1991;
Palmer, 1994), so here we only present the equation and
explain the notations. The equation can be written as
\begin{align}
 {\rm Det}\,
{\big|\!\big|}\delta^{\alpha\beta}-{\cal
M}^{\alpha\beta}(\omega){\big|\!\big|}=0\ , \ \ \alpha, \beta =
1,2,3,... \ ,
\label{me}
\end{align}
where matrix ${\cal M} ^{\alpha \beta} (\omega) $ is given by
the following expression
\begin{multline}
{\cal M}^{\alpha\beta}(\omega)=-4\pi\,G\,(2 \pi)^2\sum\limits_{l_1 = - \infty}^{\infty}
\sum\limits_{l_2 = -l}^l D_l^{l_2} \int \int \frac{dE \, dL}{\Omega_1}  \times \\
F (E, L)\,\Bigl[\Omega_{l_1l_2}\,\dfrac{\p}{\p E}+ l_2\,
\dfrac{\p}{\p L} \Bigr] \Bigl (\frac{L \, \psi ^{\alpha \beta}_{l_1
\, l_2}}{\omega- \Omega_{l_1 l_2}} \Bigr)\ . \label{eq:lme}
\end{multline}
The integration in (\ref{eq:lme}) is taken over the allowed
domain ${\cal D} $ of two-dimensional action sub-space $ (E, L) $.
This is the so-called `Lagrangian' form of the matrix elements
(see Appendix A). It is different from the more familiar
`Euler' form:
\begin{multline}
{\cal M} ^{\alpha \beta} (\omega)=4\pi\,G\,(2 \pi)^2\sl_{l_1=-\infty}^{\infty}
\sum\limits_{l_2 = -l}^l D_l^{l_2} \int \int \frac{dE L dL}{\Omega_1 (E, L)} \times \\
\psi ^{\alpha \beta}_{l_1\,l_2}(E, L)\,\frac{\Omega_{l_1l_2} (E, L)
\, \dfrac{\p F}{\p E} + l_2 \, \dfrac{\p F}{\p L}}{\omega-
\Omega_{l_1 l_2} (E, L)} \ ,\label{eq:euler_me}
\end{multline}
which can be formally obtained by integrating by parts and
discarding the boundary terms. Note that `Euler' form
becomes incorrect for $ F (E, L) $ with an integrable singularity at
the sub-space boundary.
For our series of models, this
occurs for $q <0 $ at the boundary $ E = 0 $.

In equations (\ref{eq:lme}) and (\ref{eq:euler_me}) indeces $\alpha$
and $\beta$ correspond to the expansion of the radial part of the
potential $ \chi (r) $ and the radial part of the perturbed density
$ \Pi (r) $ over the biorthogonal set:
\begin{align}
\chi (r) &= \sum \limits_{\alpha} C ^{\alpha} \chi ^{\alpha} (r)\ , \\
\Pi (r) &= \frac{1}{4 \pi G} \sum \limits_{\beta} C ^{\beta} \rho ^{\beta} (r) \ .
\label{eq:bf_expansion}
\end{align}
The perturbations are assumed to be independent of the
azimuthal variable $ \varphi $, since the eigenfrequencies of the
perturbations $ \omega $ for spherically symmetric distributions $ F
= F (E, L) $ are independent of the azimuthal number $ m $.
Therefore, instead of the angular dependence of the general form $
\delta \Phi = \chi (r) \, Y_l ^ m (\theta, \varphi) \, e ^{- i
\omega t} $, $ \delta \rho = \Pi (r) \, Y_l ^ m (\theta, \varphi) \,
 e ^{- i \omega t} $,
one can consider simplified axisymmetric one
$ \delta \Phi = \chi (r) \, P_l (\cos \theta) \, e ^{- i \omega t} $,
$ \delta \rho = \Pi (r) \, P_l (\cos \theta) \, e ^{-i \omega t} $.
Here, functions $ \chi ^{\alpha} (r) $ and $ \rho ^{\alpha} (r) $ are related by the Poisson equation
$$
\left[\frac{d^2}{dr^2}+\frac{2}{r}\,\frac{d}{dr}-\frac{l(l+1)}{r^2}\right]\,\chi^{\alpha}=\rho^{\alpha}
$$
and satisfy the so-called biorthonormal conditions,
$$\int_0^1\chi ^{\alpha}(r)\,
\rho^{\beta}(r)\,r^2\,dr=-\delta^{\alpha \beta}\ .$$
Note that $\chi ^{\alpha} $ and $ \rho ^{\alpha} $ depend on the index $ l$,
but we omit it for brevity.

Subscripts $ l_1 $ and $ l_2 $ correspond to the decomposition
$$ \delta \Phi (I_1, I_2, w_1, w_2) = \sum \limits_{l_1 l_2}
 (\delta \Phi)_{l_1l_2} ({\bf I}) \, \exp [\, i (l_1w_1 + l_2w_2)]
$$
of the spatial dependence of the perturbed potential in harmonics
of angular variables $ w_1 $ and $ w_2 $ conjugate to action
variables $ I_1 $ and $ I_2 $:
$$
I_1 = \frac{1}{2 \pi} \oint p_r \, dr = \frac{1}{\pi} \int
\limits_{r_{\rm min}} ^{r_{\rm max}} \sqrt{2E-2 \Phi_0(r )-
\frac{L^2}{r ^ 2}}\,dr\ ,\quad I_2 = L\ .
$$
The radial angular variable $ w_1 $ is related to the radius $ r $ as follows:
$$
 w_1 = \Omega_1 \int \limits_{r_{\rm min}}^r\frac{dr'}
 {\sqrt{2E-2 \Phi_0 (r) -L ^ 2 / {r'}^2 \phantom{\big |}}}\ .
$$
An explicit expression for the angular variable $ w_2 $ can be found
in the mentioned above papers (e.g., Polyachenko \& Shukhman, 1981). For the perturbations
independent of the azimuthal variable $ \varphi $,  $ \delta \Phi $
and $ \delta \rho $ do not depend on the angular variable $ w_3 $.
Functions $ \Omega_{l_1l_2}(E, L)$ denote linear combinations
of orbital frequencies, $ \Omega_{l_1l_2} \equiv l_1 \Omega_1 + l_2
\Omega_2 $, which are determined by:
$$
 \frac{1}{\Omega_1} = \frac{1}{\pi} \int \limits_{r_{\rm min}}^{r_{\rm max}}
 \frac{dr}{\sqrt{2E-2\Phi_0(r)-L^2/r^2\phantom{\big |}}}\ ,
$$
and,
$$
\frac{\Omega_2}{\Omega_1} = \frac{L}{\pi} \int \limits_{r_{\rm
min}}^{r_{\rm max}} \frac{dr}{r ^ 2 \, \sqrt{2E-2 \Phi_0 (r)-L^2/r^2
\phantom{\big |}}} \equiv \frac{\Delta \varphi}{\pi}\ ,
$$
where $ \Delta \varphi (E, L)$ is the angular distance in the orbital
plane between $ r_{\rm min} $ and $ r_{\rm max} $. For highly
eccentric orbits and non-singular unperturbed potentials -- the cases in which we are
interested in, this angle is close to $ \frac{1}{2} \, \pi $,
giving the frequency ratio $ \Omega_2 / \Omega_1 \approx \frac{1}{2}
$. Such orbits, called 2:1-orbits, are slowly precessing
ellipses symmetric relative to the centre.

The coefficients $ D_l^k $ are nonzero only for even $|l-k|$ and
equal
$$
 D_l ^ k = \dfrac{1}{2 ^{2\,l}}\,\dfrac{(l+k )!(l-k)!}
 {\Bigl[\bigl(\frac{1}{2}\,(l-k)\bigr)!\,\bigl(\frac{1}{2}\,(l+k)\bigr)!{\phantom{\big|}}\Bigr]^2}\ .
$$
 Finally, $ \psi ^{\alpha \beta}_{l_1 \, l_2} (E, L)
 = \phi ^{\alpha}_{l_1l_2} (E, L) \phi ^{\beta}_{l_1l_2} (E, L)$,
 where,
$$
 \phi_{l_1 \, l_2} ^{\alpha} (E, L) =
 \frac{1}{\pi} \int \limits_0 ^{\pi} \cos \Theta_{l_1 l_2} (E, L; w_1) \,
 \chi ^{\alpha} \bigl [r (E, L, w_1) \bigr] \, dw_1\ ,
$$
 and the angle
 $ \Theta_{l_1l_2} (E, L, w_1) $ is defined by $ \Theta_{l_1 \, l_2} (E, L; w_1)
  = \Omega_{l_1l_2} \, \dfrac{w_1}{\Omega_1} -l_2 \delta \varphi (E, L; w_1), $ where
$$
 \delta \varphi (E, L, w_1) = L \int \limits_{r_{\rm min} (E, \, L)} ^{r (E, L, w_1)}
  \frac{dx}{x \, \sqrt{\phantom{\big |} [2E + 2 \Psi (x)] \, x ^ 2-L ^ 2}}
$$
is the angular distance between $ r_{\rm min} $ and the
current radius $r$; the relative potential $ \Psi (r) \equiv - \Phi_0 (r)> 0 $.

\section{Results}

Equilibrium models of $ q = 0 $ and $ q = -1 $ were
analysed in detail in Polyachenko et al.
(2013). In both cases, limiting models $ L_T = 0 $ describe
systems of purely radial orbits with global anisotropy $ \xi \equiv
1- \zeta ^{- 1} = 1 $. Nearly radial models corresponding to small $
L_T $ have isotropic and almost homogeneous kernel with radius $ r_1
\sim L_T $. Within this radius, the potential is almost constant: $
\Psi (r) = \Psi (0) +{\cal O} (r ^ 2) $, $\Psi (0)> 0 $. For $ r >
r_1 $, the density and the potential vary as follows:
$$
 \rho \sim \frac{\Psi ^{q + 1/2}}{r ^ 2} \ , \quad \Psi
 \sim \ln ^ n (1 / r) \ , \quad n = (1/2-q) ^{- 1} \ .
$$

\subsection{$q=0$ series}

\begin{figure}
\centerline{\includegraphics [width = 85mm]{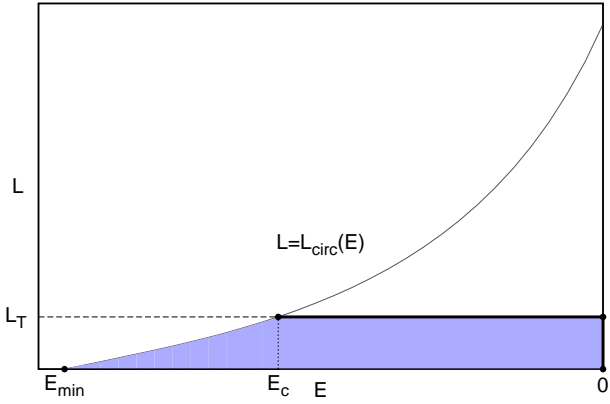}}
\caption{$(E,L)$ sub-space and the allowed domain ${\cal D}$
(filled area). The upper curve is the line of circular orbits.}
\label{EL}
\end{figure}

The DFs $ F (E, L) $ in $ q=0 $ series are independent
of the energy $ E $ and the angular momentum $L$ within the
allowed domain $ \cal D $:
\begin{align}
 F (E, L) = \frac{N (0,L_T)}{2 \pi ^ 3} \, \frac{H (L_T-L)}{L_T ^ 2} \, H (-2E ).
\label{F0}
\end{align}
For the chosen form of DF, the expression for matrix elements
$ M ^{\alpha \beta} (\omega) $ is particularly simple, since the
two-dimensional integration over ${\cal D} $ is reduced to
one-dimensional integration along two boundary lines: the vertical $
0 <L <L_T$, $ E = 0 $, and the horizontal $ L = L_T$, $E_c <E <0$
(shown by thick lines in Fig. \ref{EL}).
The models become isotropic when the parameter $ L_T \geq 0.6682 $ (Polyachenko
et al., 2013).

A suitable expression for the matrix elements $ M ^{\alpha
\beta} (\omega)$ can be obtained from (\ref{eq:euler_me}) and
written in a form containing eigenfrequency squared and summation
over non-negative $ l_1 $ only:
$$
 M ^{\alpha \beta} (\omega) = \frac{8N}{L_T ^ 2}
 \int \limits_{0} ^{L_T} L dL \,{\cal S} ^{\alpha \beta}(0, L)-
 \frac{16N}{L_T} \sum \limits_{l_1 = 0} ^{\infty} S_{l_1} \sum \limits_{l_2 = -l} ^ l D_l ^{l_2} \times
$$
$$
\times \left\{\frac{1}{L_T} \int \limits_0 ^{L_T}{L \, dL} \,
 \frac{\omega ^ 2 \, \psi ^{\alpha \beta}_{l_1 \, l_2} (0, L)}{\Omega_1 (0, L) \,
 [\, \omega ^ 2- \Omega_{l_1l_2} ^ 2 (0, L)]} \, + \right.
$$
\begin{align}
  +\left. \int\limits_{E_c} ^ 0 dE \, \frac{\Omega_{l_1 l_2} (E, L_T) \,
   \psi ^{\alpha \beta}_{l_1 \, l_2} (E, L_T)}{\Omega_1 (E, L_T) \,
    [\, \omega ^ 2- \Omega_{l_1l_2} ^ 2 (E, L_T)]} \right \}\ .
\label{me_q = 0}
\end{align}
Here, the coefficients $S_{l_1} = 1/2$
for $l_1 = 0 $, and $S_{l_1} = 1$ otherwise.
The first term on the r.h.s. of (\ref{me_q = 0})
results from a summation of terms independent of $ \omega ^ 2 $:
$$
{\cal S} ^{\alpha \beta} (E, L) = \frac{2}{\Omega_1 (E, L)}
\sum \limits_{l_1 = 0} ^{\infty} S_{l_1}
\sum \limits_{l_2 = -l} ^ l D_l ^{l_2} \psi ^{\alpha \beta}_{l_1l_2} (E, L),
$$
which can be performed analytically (see, e.g., Saha 1991):
\begin{align}
{\cal S} ^{\alpha \beta} (E, L) = \frac{1}{\pi} \int_{r_{\rm min}} ^{r_{\rm max}}
\frac{\chi ^{\alpha} (r) \, \chi ^{\beta} (r)} {[2E-2 \Phi_0 (r) -{L ^ 2} /{r ^ 2}] ^{1/2}} dr \ .
\label{cal_S}
\end{align}

Using equations (\ref{me}) with matrix elements (\ref{me_q = 0}), we
investigate the stability of spherical harmonics $ l $ in the range
$ 1 \leq l \leq $ 20 for $ L_T \geq 0.01 $. The model $ L_T =
0.01 $ is highly radially anisotropic with ratio $ \zeta \equiv 2T_r
/ T_ \perp = $ 42, and the global anisotropy $ \xi = 0.98 $.

Success of the matrix method depends largely on the appropriate
choice of basis functions $ \{\chi ^ \alpha, \rho ^ \alpha \} $. In
Appendix B  we describe a method for constructing a variety
of basis sets in which $ \rho^\alpha=\lambda^\alpha
g(r)\,\chi^\alpha $, where function $g(r)$ is arbitrary. The special
case $ g (r) = -1 $ corresponds to a well-known orthogonal system of
spherical Bessel functions (Polyachenko and Shukhman, 1981).

A numerical code for mode's calculation was tested by finding
lopsided $l=1$ shear modes, which is present in all models. They are
called \textit{zero modes}, since its eigenfrequency $ \omega = 0 $,
and the eigenfunction $ \chi = \chi ^{1} (r) = A \Psi'$.  Recall
that shift $ \delta z $ of the sphere as a whole along $ z $-axis
generates a perturbation of the potential $ \delta \Phi = \delta z
\cdot  \cos \theta \, \Phi_0'\equiv \delta z \,\Phi_0'\,P_1(\cos
\theta)$. For small $\omega^2$, the matrix elements can be represented as a series in
$\omega^2$:
\begin{align}
{\cal M} ^{\alpha \beta}(\omega^2)=a^{\alpha \beta}+b^{\alpha \beta}
\omega^2+{\cal O}(\omega^4)\ ,
\label{eq:msw}
\end{align}
 and hence ${\rm Det} \, \|{\cal M} ^{\alpha \beta}(\omega^2)-
 \delta_{\alpha \beta}\| \approx A + B \, \omega ^ 2$.
For the zero modes, $ A $ must vanish. In fact, it is not zero
due to different approximations, such as, using a grid in $(E, L)$
sub-space, substitution of the infinite matrix $ M ^{\alpha \beta} $
by a matrix of finite size ($ N_{\alpha} \times N_{\alpha}) $, and
the replacement of an infinite series in $ l_1 $ by finite series of
length $ (l_1)_{\rm max} $. Assuming that an error is $ \delta
\omega \equiv | A / B | ^{1/2} $, we can select the best basis by
testing different functions $g(r)$. It turns out that for $ L_T =
0.1 $ the appropriate choice is $g(r)=-\rho(r)$, with $ \delta
\omega \approx 0.03$. The zero mode test allowed us to verify the
accuracy of the equations and obtain the accuracy estimate
of eigenfrequency calculations. We also note that the accuracy
drops sharply at $ L_T <0.01 $.\footnote{For very small $L_T$, we
have developed recently a special approach which allows to
investigate spectrum of eigenfrequencies even for almost pure radial
models, $L_T\to 0$. This approach and its application will be
presented in separate work.}

For $ g (r) = \rho'_0(r)/\Phi'_0(r)$, the shear mode eigenfunction
for the potential consists of only one element: $ \chi ^{\alpha} (r)
= A \, \Psi'(r) \, \delta ^{1, \alpha} $. So, all the elements
in the first row and first column of matrix ${\cal M} ^{\alpha
\beta} (0) $ must be close to zero, except that the first one is equal to
unity.

\begin{figure}
\centerline{ \includegraphics [width = 85mm, draft = false]{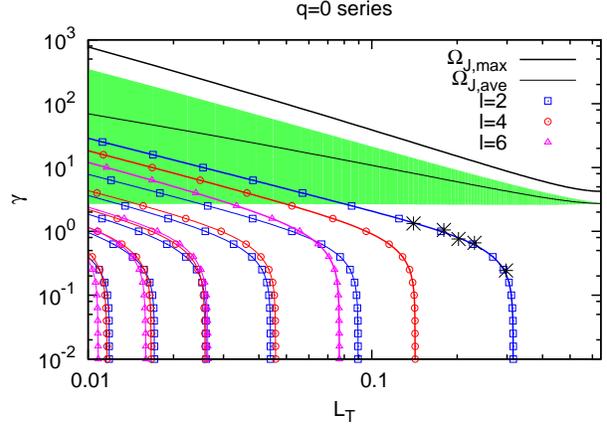}}
\caption{The dependence of the growth rate $ \gamma $
of aperiodic modes on the parameter $ L_T $ for $ l = 2, \, 4, \, 6$. Stabilization of the most unstable mode $ l = 2 $ occurs at $ L_ \textrm{crit} = 0.316 $. Profiles of characteristic frequencies are given by (\ref{eq_O1_max}). The shaded area shows the range of variation of the radial frequency $ \Omega_1 $, $(\Omega_J)_{\textrm{LOC}} $, which almost coincides with the maximum radial frequency $ \Omega_1 $ (not shown). Results of the growth rate $N$-body calculation for $l=2$ are given by asterisks.}
\label{eigen_q = 0}
\end{figure}

The results of the stability study are the following.

1. There are no unstable solutions corresponding
to odd spherical harmonics $ l $.

2. For even values of $ l $ we found only {\it aperiodic}
unstable solutions, ${\rm Re} \, \omega = 0 $, $ \omega = i
\gamma$.

3. The unstable models are found within the range $ L_T <
0.316 $, which corresponds to $ \zeta \equiv 2 \, T_r / T_{\perp}>
2.2 $, or global anisotropy $ \xi> 0.55 $.

4. For a given spherical harmonic $ l $, the number of unstable
modes increases infinitely with decreasing $ L_T $.
These modes have different growth rates
$ \gamma_j ^{(l)} (L_T) $,  $ j = 1,2, ..., j_{\rm max} ^{(l)} (L_T) $
 (see Fig.\,\ref{eigen_q = 0}).

\begin{figure}
\centerline{ \includegraphics[bb=50 50 266 192,clip ,width=85mm, draft=false]{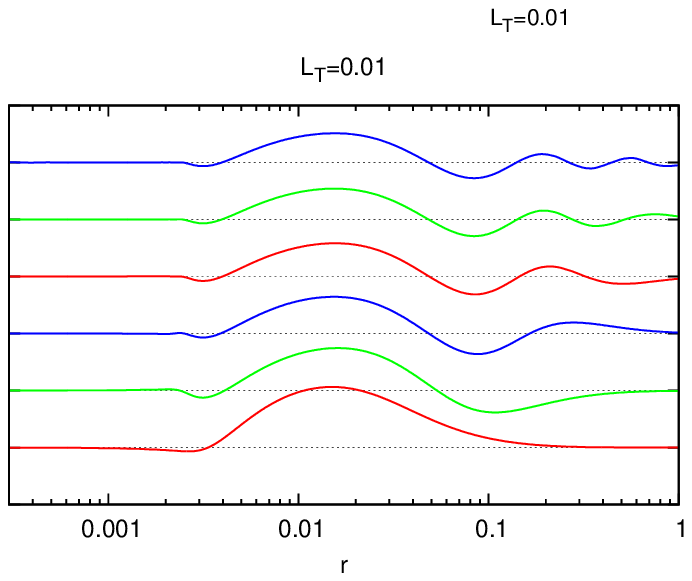}}
\caption{Radial profiles of the potential perturbation for the model $ L_T = 0.01 $
corresponding to different modes of unstable spherical harmonic $ l = 2 $.
To save room the profiles are shown in
a single plot, but vertically separated from one another ($\gamma_j$ decreases from bottom to top).}
\label{fig_modes_q0}
\end{figure}

5. Eigenfunctions of the radial part of the potential perturbation,
corresponding to different modes $j$ differ in the number of
nodes (see Fig.\,\ref{fig_modes_q0}). Larger growth rates $ \gamma_j
$ correspond to modes with fewer nodes.

6. With increasing $ L_T $, the growth rates $ \gamma_j ^{(l)} $
decrease, and modes with large number of nodes disappear.

Stabilization of all modes of $ l = 4 $ harmonic occurs at $ L_T =
0.142 $ ($ \zeta = 5.1 $; $ \xi = 0.8 $); all $l = 6$ modes
stabilize at $ L_T = 0.077 $ ($ \zeta = 8.7 $; $ \xi = 0.89 $). The
largest-scale nodeless mode ($l=2$) remains the most unstable for any
value of parameter $ L_T < 0.316 $.

Fig.\,\ref{eigen_q = 0} compares
the obtained growth rates of $ l = 2, 4, 6$ unstable modes, and some
characteristic frequencies, such as radial frequency oscillations $
\Omega_1 $, and
\begin{align}
 (\Omega_J)_{\textrm{max}} &= \sqrt{4 \pi G \rho (0)}\ , \nonumber \\
 (\Omega_J)_{\textrm{ave}} &= \sqrt{4 \pi G \int dr \, r ^ 2 \rho ^ 2 (r) / M}\ , \label{eq_O1_max} \\
 (\Omega_J)_{\textrm{LOC}} &= \sqrt{4 \pi G \rho (2r_1)}\ , \nonumber
\end{align}
where $ r_1 $ is the isotropic radius of the nucleus. $
(\Omega_J)_{\textrm{ave}} $ is a weighted Jeans frequency; $
(\Omega_J)_{\textrm{LOC}} $ is a Jeans frequency on the radius of
localization of the perturbation (i.e., $ r_{\rm LOC} \approx 2 \,
r_1 $).
It is seen that
the growth rates are small compared with all characteristic
frequencies (\ref{eq_O1_max}): for instance, $
(\Omega_J)_{\textrm{LOC}} $ exceeds the growth rates for more than
an order of magnitude.

The obtained slowness of the modes allows one to average
over the motion of a particle along its orbit, and consider orbital
slow dynamics rather than ordinary particle dynamics (see, e.g.,
Polyachenko, 2004, 2005). Then the matrix equation can be
simplified, and instead of the full equation based on matrix
(\ref{me_q = 0}), one may consider a `slow' equation, which is
obtained from (\ref{me_q = 0}) by omitting all terms except those
for which $ l_1 = -l_2 / 2 $. Similar simplification is used by Palmer (1994) for calculation and interpretation of the instability in the `Lynden-Bell' approach. Recall that for highly eccentric
orbits $ \Omega_2 \approx \Omega_1 / 2 $, which means
\begin{align}
 l_1 \Omega_1 + l_2 \Omega_2 = l_2 \, (\Omega_2 -{\case{1}{2}}\,
  \Omega_1) = l_2 \, \Omega_{\rm pr} \ll \Omega_{1,2}.
\label{eq_res}
\end{align}
We checked the applicability of the `slow' approach by direct
 recalculation of the spectra of modes for the lower spherical
harmonics ($l\leq 6$). The comparison demonstrates
that difference in frequency values does not usually exceeds 1 per cent.
Thus we conclude that for $q=0$ series ROI can be interpreted in terms of `Lynden-Bell' mechanism.

This result seems suspicious
in the absence of a dominant external potential, which provides a
slow precession for all orbits and ensures the slow mode
(Polyachenko et al., 2010). An order-of-magnitude estimate for
eigenfrequencies gives the Jeans frequency, or the dynamical
frequency, $ \omega \sim \Omega_J \sim \Omega_{\rm dyn} \sim
\sqrt{GM / R ^ 3} = 1 $.
However, this is a hasty conclusion: although the
characteristic dynamical frequency is of the order unity, the
maximum dynamical frequency of star oscillations `locked' near the
centre is very high. For $q = 0$ series a large group of stars never
leaves the centre, and despite the obtained growth rates are
substantially greater than unity, they are still much smaller than
the dynamical frequency of the locked stars. The slowness occurs here due to the small deviation of the potential from a harmonic form that exists in the central region. Note that these modes are turn out to be analogous to slow modes in near-Keplerian systems (Tremaine, 2001): in both cases orbital precession rates are low, and the modes are formed due to orbit--orbit alignment.

We performed an `experiment' to determine which particles give
the main contribution to the growth rates, retaining only
contribution from $ E_c $ to some $ E = E_{\textrm{max}} <0 $ in the
integrals over the horizontal line $ L = L_T, \ E_c <E <0 $. Our
calculations confirm that the main contribution to the matrix
elements comes from particles with apocentric distances $ r_{\rm
max} (E, L = L_T) $ much smaller than unity.

Jeans instability mechanism suggests a simple dependence of
critical parameter $(L_T)_{\rm crit}$, at which the system becomes
stable, from the spherical harmonics $ l $.
For a system in equilibrium $ \sigma_r \sim R \, \sqrt{G
\rho} $, where $ \sigma_r $ is the radial dispersion. On the other
hand, assuming marginal stability one can obtain for the Jeans
characteristic scale in the transverse direction, $ \lambda_J $,
from the Jeans criterion $ \lambda_J \sim \sigma_{\perp} / \sqrt{G
\, \rho} $. Thus, using $ \lambda_J \sim R / l $, we have
\begin{align}
 \frac{1}{l_ \textrm{crit}} \sim \frac{\lambda_J}{R} \sim \frac{\sigma_ \perp}{\sigma_r}\ ,
\end{align}
which gives $ l_ \textrm{crit} \sim (1- \xi) ^{- p} $ with the exponent $ p $ of order unity.
Since for small $L_T$, $L_T \simeq (1-\xi)(1/2-q)$ (Polyachenko et al., 2013),
it is natural to expect $ (L_T)_{\rm crit} $ to be
inversely proportional to some power of $ l $: $(L_T)_{\rm crit}  \propto l^{-1/p}$.

\begin{figure}
\centerline{
\includegraphics [width = 85mm]{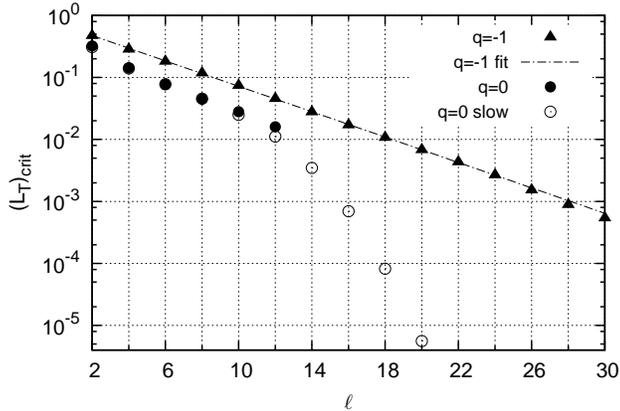}}
\caption{The dependence of the critical parameter $ (L_T)_{\rm
crit}$ on the harmonic number $ l $: $ q = 0 $ series (filled
circles show the stability boundary inferred from the full integral
equation (\ref{me_q = 0}), open circles from the `slow' integral
equation);  $ q = -1 $ series is shown by triangles (full equation).
The dash-dotted line shows the least square linear fit to the $ q =
-1 $ boundaries.} \label{L_T_crit}
\end{figure}

Fig.\,\ref{L_T_crit} shows the stability boundaries $ (L_T)_{\rm
crit} $ for even harmonic numbers $ l $ in the range $ 2 \le l \le $
30. The filled circles show the results obtained using the full
matrix equation (\ref{me_q = 0}). Starting from $ l = 10 $, the
linear combination of orbital frequencies $ \Omega_{l_1l_2} \equiv
l_1 \Omega_1 + l_2 \Omega_2 $ vanishes for some $ l_1, l_2 $. Due to
these resonances, calculation of the stability boundaries becomes
extremely time-consuming. Open circles show the results of
calculations using the `slow' equation, which are almost identical
in the absence of resonances ($ l \leq $ 8). However, when
resonances appear, the results begin to diverge. The full equation
gives approximately exponential decay for $ (L_T)_{\rm crit} (l) $,
while the `slow' solution decreases significantly faster than
exponent.

We conclude that the estimate for $ (L_T)_{\rm crit} (l) $
based on usual Jeans relations for gravitating medium is
incorrect to describe the bar-forming instability in highly
heterogeneous systems. It enables only to predict the decrease with $ l $.
In addition, `slow' solution is applicable only in the absence of resonances.
Correct calculations of spherical harmonics $ l \geq 10 $
are possible by using the full equation only.

Our matrix calculations was also supported by numerical
$N$-body simulations using Superbox-10 code (Bien et al.
2013). This code is an example of a particle\,--\,mesh scheme, which
solves the Poisson equation by fast Fourier transform. The number of
grid points $ N_g $ for each coordinate is the same and is taken so
that the number of grid cells be comparable with the number of
particles. The number of particles in all calculations except one
was $ 10^6 $; with $ N_g = 256$. The model close to the
stability limit ($ L_T = 0.2975 $) was calculated with $10^7$ particles and
$ N_g = $ 512. The code uses three meshes. The biggest one allows us to
simulate interaction between galaxies. Medium meshes are designed to
simulate separate galaxies, and the smallest meshes are used to
resolve fine structures in galactic centres. In all our
calculations, the mesh sizes were taken to be 30, 5 and 1 (recall
that the initial radius of the system $R=1$). The growth
 rates, evaluated from numerical experiments show good
 agreement with results of matrix calculations,
 especially for the models with moderate growth rates.
 On the other hand, in the models with $ L_T < 0.14 $
 it is difficult to distinguish the temporal interval of exponential
 growth of perturbations. This can be explained by
 interference of $ l = 2 $ and $ l = 4 $ modes.
 To investigate these models, initial states of high degree
 of symmetry as well as filtering of higher harmonics of force are needed.

\subsection{$q=-1$ series}

In contrast to $q=0$ series discussed above, models of $q=-1$ series are
monoenergetic, e.g. all stars have the same energy. In the limit of
purely radial orbits all of them reach the outer radius; thus stars
locked near the centre are absent. Accordingly, the dynamical
frequencies of  stars are almost identical and are of the order
 unity. Here we investigate how this affects the
stability properties.

For the given $q$, the distribution function (\ref{apm}) reduces to:
\begin{align}
 F(E, L) = \frac{N (-1, L_T)}{4 \pi ^ 3} \, \frac{H (L_T-L)}{L_T ^ 2} \, \delta (E),
\label{F-1}
\end{align}
where $ \delta (x) $ is the Dirac $ \delta
$-function. In the limiting case of purely radial orbits, $
L_T = 0 $, we obtain a model discussed by Agekyan (1962). On
the other hand, systems become isotropic at $ L_T \geq 0.5613 $.

In the calculation of the matrix element, $ M ^{\alpha \beta}
(\omega) $, two-dimensional integration is reduced to a one-dimensional
 integration over the vertical interval $ E = 0, \ 0 <L \le L_T $.
 However, the integrand includes a derivative with respect to energy,
 so some functions should be found in a small neighbourhood
 $ \Delta E <E <0, 0 <L <L_{\rm circ} (E) $.
 The final expression of the matrix
 element obtained from (\ref{eq:euler_me}) has a form:
$$
 M ^{\alpha \beta} (\omega) = \frac{2{N}}{L_T^2}\,\frac{d}{dE}
  \left[\int \limits_0 ^{L_T} L dL \,{\cal S} ^{\alpha \beta} (E, L) \right]_{E = 0} -
\frac{4{N}}{L_T} \sum \limits_{l_1 = 0} ^{\infty} S_{l_1} \times
$$
$$
 \sum \limits_{l_2=-l}^l D_l^{l_2}\left\{\left[\frac{d}{dE}\int\limits_0^{L_T}\frac{L\,dL}{L_T}\,
 \frac{\omega^2\,\psi^{\alpha\beta}_{l_1\,l_2}(E,L)}{\Omega_1(E,L)\,
 [\,\omega^2-\Omega_{l_1 l_2} ^ 2 (E, L)]} \, \right]_{E = 0} + \right.
$$
\begin{align}
 \left. + \frac{\Omega_{l_1 l_2} (0, L_T) \,
 \psi ^{\alpha \beta}_{l_1 \, l_2} (0, L_T)}{\Omega_1 (0, L_T) \,
 [\, \omega ^ 2- \Omega_{l_1 l_2} ^ 2 (0, L_T)]} \right \}\ ,
\label{me_q = -1}
\end{align}
 where
 ${\cal S} ^{\alpha \beta} (E, L) $
 is defined by (\ref{cal_S}).
 The first term on the r.h.s. of (\ref{me_q = -1})
 can be converted to
$$
\frac{2{N}}{L_T ^ 2} \, \frac{d}{dE} \left[\int \limits_0 ^{L_T} L
dL\,{\cal S} ^{\alpha \beta} (E, L) \right]_{E = 0} = -
\frac{2N}{\pi \, L_T ^ 2} \times
$$
\begin{align}
 \left[\int \limits_{r_1} ^{r_2} \frac{\chi ^{\alpha} (r)\,\chi^{\beta}(r)\,r^2\,dr}
 {\sqrt{2 \Psi(r)-L_T^2/r^2}}-\int\limits_0^1\frac{\chi^{\alpha} (r)\,\chi^{\beta}(r)\,r^2\,dr}
 {\sqrt{2 \Psi (r)}} \right],
\label{me_q = -1A}
\end{align}
where in the first integral on the r.h.s. of (\ref{me_q = -1A})
$r_{1,2}$ are zeros of the radicand in the denominator.

\begin{figure}
\centerline{
\includegraphics [width = 85mm]{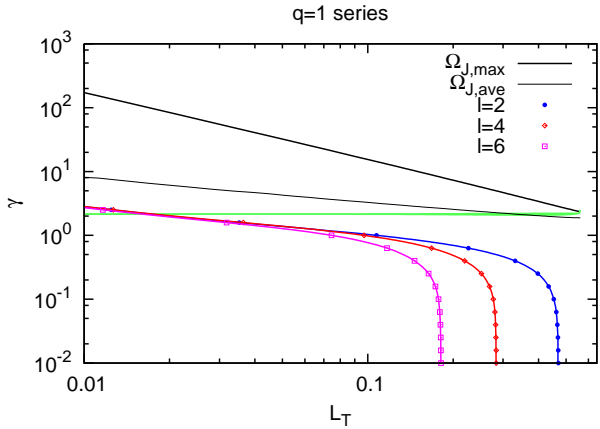}}
\caption{Same as in Fig.\,\ref{eigen_q = 0} for $q=-1$ series.
Stabilization of the most unstable mode $ l = 2 $ occurs at $ L_
\textrm{crit} \approx 0.47 $. The shaded area of  $ \Omega_1 $
variation is degenerated into a thin almost horizontal line.}
\label{even q = m1}
\end{figure}

Stability of individual harmonics is investigated
by employing equation (\ref{me}) with the matrix elements (\ref{me_q = -1})
for $ 1 \le l \le $ 30 and $ L_T \gtrsim 10 ^{- 3} $.
Note that for the model with $ L_T = 0.01 $, parameter
$ \zeta = 2 \, T_r / T_{\perp} = $ 172
and the global anisotropy $ \xi = 0.994 $.

The spectrum of unstable modes for models of this series
differs significantly from the spectrum of $ q = 0 $ series models.
This applies to both even and odd spherical harmonics $ l $.

\begin{figure}
\centerline{\includegraphics [width = 85mm]{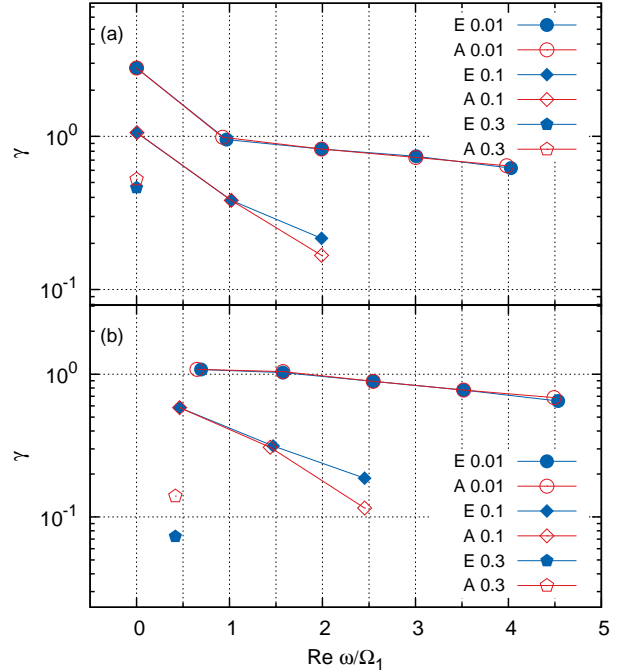}}
\caption{The oscillating modes of $ q = -1 $ series for three values of $ L_T $ ($ L_T = 0.01; \, 0.1; \, 0.3 $): (a) $ l = 2 $, (b) $ l = 3 $. Solution (E) obtained by exact matrix equation (\ref{me_q = -1}); solution (A) by an approximate equation (see explanation in the text). All frequencies are given in units of the radial frequency for purely radial orbits $ \Omega_1 \approx 2.16 $.}
\label{q1cr}
\end{figure}

1. For even $ l $, there is only one aperiodic unstable mode,
$$
 \omega ^{(l)}_ 0 (L_T) = i \gamma_0 ^{(l)} (L_T)
$$
 and several oscillating unstable modes
$$
 \omega ^{(l)}_ j (L_T) ={\bar \omega}^{(l)}_j(L_T)+i\gamma_j ^{(l)} (L_T), \ j = 1,. .., j_{\rm max},
$$
in which the growth rates decrease with real part of the frequency increasing,
$$
{\bar \omega}_{j-1} ^{(l)} (L_T) <{\bar \omega}_j ^{(l)} (L_T), \\\gamma_{j-1} ^{(l)} (L_T)> \gamma_j ^{(l)} (L_T),
$$
 and the number of unstable oscillatory modes $ j_{\rm max} $ depends on $ L_T $. The real parts of the frequencies are separated by $\Omega_1$ (see Fig.\,\ref{q1cr} a):
$$
{\bar \omega} ^{(l)}_{j} -{\bar \omega} ^{(l)}_{j-1} \approx \Omega_1 (E = 0, L = L_T).
$$

2. For odd modes starting with $ l = 1 $, there are only
oscillating unstable modes, the real part of the frequencies are approximately equally spaced (see Fig.\,\ref{q1cr} b).

3. Growth rates of all modes decrease with increasing $ L_T $.
For a given even spherical harmonics $ l $,
less unstable oscillatory modes stabilize first,
then more unstable aperiodic modes stabilize.

\medskip

All non-spherical harmonics are fully stabilized when $ L_T> (L_T)_{\rm crit}
\approx 0.47 $, corresponding to $ \zeta <1.5 $ or $ \xi <0.34 $.
It coincides with the stabilization of aperiodic bar-mode instability,
$ l = 2 $ (see Fig. \ref{even q = m1}).
Stabilization of $ l = 4 $ harmonic occurs
at $ L_T = 0.284 $ ($ \zeta = 3.7 $; $ \xi = 0.73 $)
and $ l = 6$  harmonic -- at $ L_T = 0.182 $ ($ \zeta = 6.9 $; $ \xi = 0.85 $).

Fig.\,\ref{L_T_crit} shows the stability boundary $ (L_T)_{\rm crit}
$ for even modes in the range  $ 2 \le l \le $ 30. The unstable
modes (shown by triangles) are
obtained using the full matrix equation (\ref{me_q = -1}). They fit
well a simple relation:
$$
 \ln (L_T)_{\rm crit} = - 0.2359 \, l - 0.2717\ .
$$
Slight deviation from linearity at $ l \geq $ 26 may be due to insufficient accuracy of the calculations.

Fig.\,\ref{q1cr} shows oscillatory and aperiodic modes of even
harmonics for different values of $ L_T $. Closed symbols show the
solutions obtained using the full equation (\ref{me_q = -1}),
whereas open symbols indicate eigenmodes found with a simplified
equation derived from (\ref{me_q = -1}) by neglecting the terms
associated with the energy derivative of the DF. The remaining last
term in (\ref{me_q = -1}) mostly determines the eigenfrequencies, which
also follows from the figure. For strongly radially anisotropic
models, the growth rates of aperiodic modes are much larger than
unity. However, at $ L_T \gtrsim 0.1 $ they become of the order
unity and comparable with the growth rates of oscillatory modes.
Note that the radial frequency $\Omega_1(E,L) $ enters the equation
at the boundary term $E= 0$, $L \leq L_T $ only, and for small $L_T$ is close to a limit $\Omega_1 (0,0) \approx 2.16$. Thus, all
the modes in this series can not be considered as slow ones.

Note that in contrast to $q=0$ models, in which some stars are locked in the centre, in $q=-1$ models the orbits of stars have large radial excursions and are no longer in the regime of slow dynamics. This is a possible reason for the fastness of the $q=-1$ modes.

\section{Conclusion}

This paper analyses two approaches presented in the literature to
interpretation of the radial orbit instability (ROI). The first one
explains ROI in terms of the classical
Jeans instability. Indeed, it is natural to expect that radially anisotropic
systems are `cold' enough in the transverse direction for the instability to develop.
Such an approach has been proposed in the first works
devoted to ROI (Polyachenko and Shukhman 1972, 1981).
The second approach appeals to a bar formation in disc galaxies proposed by Lynden-Bell (1979). His mechanism considers
coalescence of the so-called `abnormal' stellar orbits, for which the precession rates of stars, $ \Omega_{\rm pr} $, decreases with decrease of angular momentum $ L $  while the adiabatic invariant $ J_f = I_R + \frac{1}{2} \, L $ is conserved (here $ I_R $ is the radial action), i.e. $(\p\Omega_{\rm pr}/\p L)_{J_f}>0$.
Let a weak bar-like perturbation of the potential rotates at a slow rate $ \Omega_p $. Orbits having a form of nearly symmetric ovals 
and the precession rates close to bar pattern speed effectively interact with the bar. The torque exerted by the bar on the abnormal orbit results in a change of the orbit precession rate, so that the orbit tends to line up with the bar, contributing to the potential well, thus enhancing the bar-like perturbations.

For spherical systems, highly eccentric orbits usually obey
required inequality ${\p \Omega_{\rm pr}}/{\p L}>0$.
Therefore, the idea of Lynden-Bell can allegedly be extended to
spherical systems. However, the key assumption that makes this
analogy legitimate is {\it slowness} of the perturbations, i.e.
perturbation frequencies must be much smaller than the
characteristic orbital frequencies. Otherwise, the concept of
orbit as a separate united object (instead of a set of individual
stars) interacting with potential perturbation is invalid.
Moreover, the orbital approach can be used to describe symmetric relative to the center
disturbances only. The easiest way to explain it is to consider
perturbations on a disc, which have a form $ \delta
\Phi \propto \cos (m \varphi)$. Even $m$ describe symmetric
perturbations relative to the center, while for odd $ m $ signs of
the potential are opposite at the opposite points: $\Phi (r,
\varphi+\pi)=-\Phi(r, \varphi)$. In the latter case, the potential
exerts differently on either side of the symmetric elliptic orbit.
Note that for spherical models, the role of the azimuthal
number $ m $ plays a spherical number $ l $.
If an unstable model violates any of these assumptions (the
perturbation is not slow, or instability is possible for odd $ l $),
then the validity of the `Lynden-Bell' or, equivalently, `slow even-$l$' or `orbital') interpretation for this model
can be put into question.

In this paper, we consider two series of DFs of the form $F(E,L)$. In the first one, $ q
= 0 $, unstable perturbations are indeed slow, and are possible
for even $ l $ only. In the second series, $ q = -1 $, the modes are fast, and instability is possible both for even and odd $ l $.
The reason for this difference is in the properties of energy dependence of their DFs.  In the model with equipartition of the energy $ (q = 0) $ many stars never leave the central region (their apocentric distances are much less than the radius of the system), where frequency
of radial oscillations, $\Omega_1 $, and Jeans frequency, $ \Omega_J $, are much greater than frequencies of the perturbation. In this sense unstable modes are indeed slow, and thus the orbital approach to the instability is valid.

On the contrary, in the mono-energetic model $ (q = -1) $ orbits of
all stars nearly approach the outer radius, so our
estimates that mode frequencies would be comparable to the
characteristic dynamical frequency (almost the same for all stars) is
proved true. Hence there is no way to use the orbital approach,
and the odd modes obeying $ | \omega | \gtrsim
\Omega_1 $ can be excited.

We conclude that the spectra of radially-anisotropic DFs of the form $F(E, L)$ are allowed to have both slow and fast modes. However, some DFs support essentially slow modes, while others allowing for both slow and fast modes.

\bigskip
\section*{Acknowledgments}
The authors thank the referee for providing several valuable suggestions for
presentation of the material, and Dr. Jimmy Philip for editing the original
version of the article that helped to improve its quality. This work was supported by Sonderforschungsbereich SFB 881 `The
Milky Way System' (subproject A6) of the German Research Foundation
(DFG), RFBR grants No. 14-05-00080, 15-52-12387 and by  Basic Research
Program OFN-17 `The active processes in galactic and extragalactic
objects' of Department of Physical Sciences of  RAS.

\appendix

\section{Lagrangian form of matrix equation for spherical systems}

Matrix elements in equation (\ref{me}) normally has the form
\begin{multline}
M^{\alpha\beta}(\omega)=4\pi\,G\,(2\pi)^2
\sum\limits_{l_1=-\infty}^{\infty}\sum\limits_{l_2=-l}^l
D_l^{l_2}\int\int\frac{ dE L dL}{\Omega_1(E,L)} \\
\psi^{\alpha\beta}_{l_1\,l_2}(E,L)\,
\frac{\Omega_{l_1l_2}(E,L)\,\dfrac{\p F}{\p E}+l_2\,\dfrac{\p F}{\p
L}}{\omega-l_1\,\Omega_{l_1 l_2}(E,L)},
\label{eq:M_euler}
\end{multline}
which is sometimes called `Euler' form of the matrix elements.
The matrix contains $ E $- and $ L $-derivatives of the DF $ F (E, L) $.
Our notations are explained in the main text.

Integration is taken over domain ${\cal D} $ of the
two-dimensional sub-space $ (E,L) $, which depends on the DF.
Boundaries of  ${\cal D} $ consist of line of circular
orbits $\ell_\textrm{circ}$, line of radial
orbits $\ell_\textrm{r.o.}$, and a line of escape $\ell_\textrm{singular}$, which is usually given by $E=0$.

The Euler form of the matrix element has a
significant drawback: if the DF $ F (E, L) $ is singular on any of the boundary lines
(but the singularity is integrable), the Euler form
contains divergent integrals. Note that two natural phase
domain boundaries $\ell_\textrm{circ}$ and $\ell_\textrm{r.o.}$
are safe in this sense, because there is no flux through these lines (see below).

Our aim is to rewrite the Euler expression for the matrix
element in the form that is free of $ E $- and $ L $-derivatives
of the DF. For this we first write (\ref{eq:M_euler}) as follows
\begin{multline}
M^{\alpha\beta}(\omega)\\=4\pi\,G\,(2\pi)^2
\sum\limits_{l_1=-\infty}^{\infty}\sum\limits_{l_2=-l}^l
D_l^{l_2}\int\int\frac{ dE\,L\,dL\,
\psi^{\alpha\beta}_{l_1\,l_2}(E,L)}{\omega-l_1\Omega_1-l_2\Omega_2} \\
\times\left\{\frac{\p}{\p
E}\left[\left(l_1+l_2\,\frac{\Omega_2}{\Omega_1}\right)\!F\right]+\frac{\p}{\p
L}\left(\frac{l_2}{\Omega_1}\,F\right)\right\}\ ,
\end{multline}
using the identity
$$ \frac{\p}{\p E} \, \frac{\Omega_2}{\Omega_1} + \frac{\p}{\p L} \frac{1}{\Omega_1} = 0\ .
$$
Now we introduce a `vector' $ \bA_{l_1l_2}$ $=\Bigl ((A_E)
_{l_1l_2}, (A_L) _{l_1l_2} \Bigr) $ with coordinates
$$
(A_E) _{l_1l_2} = \left(l_1 + l_2 \, \frac{\Omega_2}{\Omega_1} \right) \! F\ , \ \ \
(A_L) _{l_1l_2} = \frac{l_2}{\Omega_1} \, F \ ,
$$
and denote
$$
B_{l_1l_2} (E, L) = \frac{L \,
\psi ^{\alpha \beta} _{l_1 \, l_2} (E, L)}{\omega-l_1 \Omega_1-l_2 \Omega_2}\ .
$$
Then the integral
\begin{multline}
    J_{l_1l_2} \equiv \int \int \frac{dE \, L \, dL \,
    \psi ^{\alpha \beta} _{l_1 \, l_2} (E, L)}{\omega-l_1 \Omega_1-l_2 \Omega_2} \\
\times\left\{\frac{\p}{\p E} \left[\left(l_1 + l_2 \,
\frac{\Omega_2}{\Omega_1} \right) \! F \right] + \frac{\p}{\p L}
\left(\frac{l_2}{\Omega_1} \, F \right) \right \}
\end{multline}
can be written in the form:
\begin{multline}
     J_{l_1l_2} = \int \int dE \, dL \, B_{l_1l_2} \,{\rm div} \bA_{l_1l_2} = \\
     \int \int \, dE \, dL \, \Bigl [{\rm div} (\bA_{l_1l_2} B_{l_1l_2}) - (\bA_{l_1l_2} \nabla) B_{l_1l_2} \Bigr]\ .
\end{multline}
According to the Gauss theorem
$$ \int \int dE \, dL \,{\rm div} (\bA_{l_1l_2} B_{l_1l_2}) = \int \limits _{\ell} d{\bell} \, (\bA_{l_1l_2} B_{l_1l_2})\ ,
$$
where $ \ell $ denotes the boundary of ${\cal D} $,
$$
\ell = \ell _{\rm circ} + \ell _{\rm r.o} + \ell _{\rm singular}\ .
$$
The DF $ F (E, L) $ is allowed to be finite or to have integrable singularity on $ \ell _{\rm singular} $.

Vector $ \bA \cdot \bB$ vanishes on the boundary $ \ell _{\rm r.o.} $, as well as the vector flux across the boundary. This is due to
the factor $ L $ in the expression for $ B_{l_1l_2} $. Furthermore, flux across the boundary $ \ell _{\rm
circ} $ vanishes due to the identity $ \phi_{l_1l_2} ^{\alpha} (E, L) \equiv \delta_{l_1 0} \chi ^{\alpha} (R) $, i.e. it vanishes on the circular orbits for all $ l_1 $ and $ l_2 $, except $ l_1 = 0 $.

Yet, flux across $\ell _{\rm singular} $ is not zero, if DF takes finite values at this boundary. Moreover, the flux is infinite if the DF is singular on $\ell _{\rm singular} $.
To summarize,
$$ \int \limits _{\ell _{\rm circ}} d{\bell} \, (\bA B) = 0\ , \ \
\int \limits _{\ell _{\rm r.o.}} d{\bell} \, (\bA B) = 0 \ , \ \
\int \limits _{\ell _{\rm singular}} d{\bell} \, (\bA B) \ne 0 \ .
$$

Thus, the Euler form of the matrix element is invalid when the DF is singular but integrable on one of the boundary lines. Besides, non-integrable singularity is present in the integral (\ref{eq:M_euler}). We may conclude that standard linearization procedure is incorrect near the boundaries if one uses the Euler technique. It is therefore clear that the attempt to bring the Euler expression to valid (Lagrangian) form without derivatives of the DF by using the Gauss theorem (roughly, using integration by parts) was bound to fail.

The desired Lagrangian form could be obtained if the flux through $\ell$ vanished. In this case, we have
$$ J_{l_1l_2} = - \int \int \, dE \, dL \, (\bA_{l_1l_2} \nabla) B_{l_1l_2},
$$
and for the matrix element:
\begin{multline}
  M ^{\alpha \beta} (\omega) = - 4 \pi \, G \, (2 \pi) ^ 2
\sum \limits_{l_1 = - \infty} ^{\infty} \sum \limits_{l_2 = -l} ^ l
D_l ^{l_2}\!\!\! \int\!\!\! \int \frac{dE \, dL}{\Omega_1} F (E, L) \\
\times\Bigl [\Omega_{l_1l_2} \, \dfrac{\p}{\p E} + l_2 \,
\dfrac{\p}{\p L} \Bigr] \Bigl (\frac{L \, \psi ^{\alpha \beta} _{l_1
\, l_2}}{\omega- \Omega_{l_1l_2}} \Bigr). \label{eq:M_lagr}
\end{multline}

However, to obtain this (correct) form, one needs to
use the Lagrangian technique from the very beginning. It was first presented in the works by Kalnajs (see, e.g. Kalnajs, 1977).
Rewriting a matrix element for disks (equation (16) of his paper)
to the spherical geometry (in which we are interested in) one can
have for the perturbations which are independent on the azimuthal angle $ \varphi $:
\begin{multline}
M ^{\alpha \beta} (\omega) = - 4 \pi \, G \,
\frac{2l + 1}{2} \sum \limits_{l_1} \sum \limits_{l_2} \int \int \int dI_1
dI_2 dI_3 F (E, L) \\
\times \Bigl (l_1 \, \dfrac{\p}{\p I_1} + l_2 \, \dfrac{\p}{\p
I_2}\Bigr)\,\left[\frac{\Psi^{\alpha\beta}_{l_1\,l_2}(I_1,I_2,I_3)}{\omega-\Omega_{l_1l_2}(I_1,I_2)}\right]\
. \label{eq:M1}
\end{multline}
Here $ I_1 $ is the radial action, $ I_2 = L $, $ \ I_3 = L_z $,
$ \Omega_{1,2} ={\p E (I_1, I_2)} /{\p I_{1,2}} $, $ \Omega_3 = 0 $.
Expression $ \Psi ^{\alpha} _{l_1 \, l_2} (I_1, I_2, I_3) $ denotes
\begin{align}
\Psi_{l_1 l_2} ^{\alpha \beta} (I_1, I_2, I_3)\!=\! \Phi ^{\alpha}
_{l_1 \, l_2} (I_1, I_2, I_3) \, [\Phi ^{\beta} _{l_1 \, l_2} (I_1,
I_2, I_3)] ^{\ast},
  \label{eq:Psi}
  \end{align}
where $ \Phi ^{\alpha} _{l_1 \, l_2} (I_1, I_2, I_3) $ can be
written in the form (see also Appendices in Polyachenko et al., 2007,
Polyachenko and Shukhman, 1981)
\begin{align}
\Phi_{l_1l_2} ^{\alpha} (E, L, \theta_0)\!=\! 2 \pi \, P_l ^{l_2}
(0) P_l ^{- l_2} (\sin \theta_0) \, e ^{il_2 \pi} \phi_{l_1l_2}
^{\alpha} (E, L)\ ,
  \label{eq:Phi}
  \end{align}
where $ \sin \theta_0 ={L_z} /{L}$, $-{\case{1}{2}} \, \pi <\theta_0 <{\case{1}{2}} \, \pi$.
From (\ref{eq:Psi}) and (\ref{eq:Phi}) one finds
\begin{multline}
\Psi_{l_1 l_2} ^{\alpha \beta} (I_1, I_2, I_3) = (2 \pi) ^ 2
P_l ^{l_2} (0) P_l ^{- l_2} (\sin \theta_0) \, e ^{il_2 \pi} \phi_{l_1l_2} ^{\alpha} (E, L)\\
\times P_l ^{- l_2} (0) P_l ^{l_2} (\sin \theta_0) \, e ^{- il_2
\pi} \phi _{- l_1, -l_2} ^{\beta} (E, L)\ .
\end{multline}
or
\begin{multline}
\Psi_{l_1
l_2} ^{\alpha \beta} (I_1, I_2, I_3) =
(2 \pi) ^ 2 \, P_l ^{l_2} (0) \, P_l ^{- l_2} (\sin \theta_0) \\
\times P_l ^{- l_2} (0) \, P_l ^{l_2} (\sin \theta_0) \,
\psi_{l_1l_2} ^{\alpha \beta} (E, L)\ .
\end{multline}
Then one needs to change the differentiation with respect to
actions $ I_1 $ and $ I_2 $ to differentiation with respect to $ E $, $ L $, $ \theta_0 $ in (\ref{eq:M1}):
$$
l_1 \, \frac{\p}{\p I_1} + l_2 \, \frac{\p}{\p I_2} = \Omega_{l_1l_2} \, \frac{\p}{\p
E} + l_2 \left[\frac{\p}{\p L} - \frac{\sin \theta_0}{L} \, \frac{\p}{\p \,
(\sin \theta_0)} \right]\ ,
$$
transform the volume element $dI_1 dI_2 dI_3={dE \, L \, dL \, d
(\sin \theta_0)} /{\Omega_1}$, and integrate over $ z~\equiv~\sin
\theta_0$. Since  $ P_l ^{l_2} (\pm 1) = 0 $ provided $ l_2 \ne 0 $,
and \[ \int _{- 1} ^ 1 dz \, z \,{d} \bigl [ P_l ^{l_2} (z) \, P_l
^{- l_2} (z) \bigr] \, dz = - \int_0 ^ 1 dz P_l ^{l_2} (z) \, P_l
^{- l_2} (z),\]
 then
\begin{multline}
\int \limits _{- 1} ^{1} dz \left(\frac{\p}{\p L} - \frac{z}{L} \, \frac{\p}{\p
z} \right) \Psi_{l_1l_2} ^{\alpha \beta} (E, L, z) \\
=\left(\frac{\p}{\p L} + \frac{1}{L} \right) \int \limits _{- 1}
^{1} dz \Psi_{l_1l_2} ^{\alpha \beta} (E, L, z)\ .
\label{eq:int_Psi}
\end{multline}
Given that the integral over $ z $ is \[\int _{- 1} ^{1} dz \, P_l
^{l_2} (0) P_l ^{- l_2} (z) P_l ^{- l_2} (0) P_l ^{l_2} (z) = [{2} /
(2l + 1)] \, D_l ^{l_2},\] one finds
\begin{align}
\int \limits _{- 1} ^{1} \Psi_{l_1 l_2} ^{\alpha \beta} (E, L, z) \, dz = (2 \pi) ^ 2
\frac{2}{2l + 1} \, D_l ^{l_2} \, \psi_{l_1l_2} ^{\alpha \beta} (E, L)\ .
\label{eq:int_Psi_1}
\end{align}
Substituting (\ref{eq:int_Psi_1}) into the r.h.s. of (\ref{eq:int_Psi}),
and then substituting the resulting expression in (\ref{eq:M1}), we obtain
\begin{multline}
M ^{\alpha \beta} (\omega) = - 4 \pi \, G \, \frac{2l + 1}{2}
\sum \limits_{l_1} \sum \limits_{l_2} \int \int \frac{dE \, L \, dL}{\Omega_1} \, F (E, L) \\
\times \left[\Omega_{l_1l_2} \, \frac{\p}{\p E} + \, l_2
\left(\frac{\p}{\p L} + \frac{1}{L} \right) \right] \frac{\int
\limits _{- 1} ^ 1 dz \Psi_{l_1 l_2} ^{\alpha \beta} (E, L,
z)}{\omega- \Omega_{l_1l_2}}\ .
\end{multline}
Finally, we have the desired expression of the matrix element in the Lagrangian form:
\begin{multline}
M ^{\alpha \beta} (\omega) = - 4 \pi \, G \, (2 \pi) ^ 2
\sum \limits_{l_1 = - \infty} ^{\infty} \sum \limits_{l_2 = -l} ^ l
D_l ^{l_2} \int \int \frac{dE \, dL}{\Omega_1}
  F (E, L) \\
 \times \Bigl (\Omega_{l_1l_2} \, \dfrac{\p}{\p
E} + l_2 \, \dfrac{\p}{\p
L} \Bigr) \Bigl (\frac{L \, \psi ^{\alpha \beta} _{l_1 \, l_2}}{\omega- \Omega_{l_1l_2}} \Bigr)\ .
    \label{eq:M_lagr_final}
\end{multline}

\section{Construction of the biortonormal basis sets}

The effectiveness of the matrix method (\ref{me}) depends on the proper choice of a basis function set $ \{\chi ^ \alpha (r), \rho ^ \alpha (r) \} $, $ \alpha = 1, 2 , ... $ satisfying the requirement of orthogonality:
\begin{align}
 \langle \chi ^{\alpha} \rho ^{\beta} \rangle \equiv \int_0 ^ 1 \chi ^{\alpha} (r) \, \rho ^{\beta} (r) \, r ^ 2 \, dr = - \delta ^{\alpha \beta}
\label{eq:ns3}
\end{align}
and the Poisson equation:
\begin{align}
\left[\frac{d ^ 2}{dr ^ 2} + \frac{2}{r} \, \frac{d}{dr} - \frac{l (l + 1)}{r ^ 2} \right] \, \chi ^{\alpha} (r) = \rho ^{\alpha} (r) \ .
\label{eq:jb}
\end{align}
 This basis is used in expansion of the perturbation potential and density (\ref{eq:bf_expansion}).

Instead of (\ref{eq:jb}), let us consider the following eigenvalue problem, assuming $ l \geq 1 $:
\begin{align}
 \left[\frac{d ^ 2}{dr ^ 2} + \frac{2}{r} \, \frac{d}{dr} - \frac{l \, (l + 1)}{r ^ 2} \right] \, \chi (r) = \lambda \, g (r ) \chi (r) \ ,
\label{eq:ns4}
\end{align}
 with boundary conditions
\begin{align}
 \chi (0) = 0 \ , \quad \chi'(1) + (l + 1 ) \chi (1) = 0 \ .
\label{eq:ns5}
\end{align}
 The solution is a discrete set of positive eigenvalues, $ \lambda ^{\alpha} $, and eigenfunctions, $ \chi ^{\alpha} (r) $, orthogonal with weight $ r ^ 2 \, g (r) $:
\begin{align}
 \int \limits_0 ^ 1 dr \, r ^ 2 g (r) \, \chi ^{\alpha} (r) \, \chi ^{\beta} (r) \propto \delta ^{\alpha \beta}\ .
\label{eq:ns7}
\end{align}
Then the biortonormal set consists of functions $ \chi ^{\alpha} (r) $ for the potential and functions $ \rho ^{\alpha} = \lambda ^{\alpha} \, g (r) \, \chi ^{\alpha} (r) $ for the density. According to (\ref{eq:ns7}), the biortonormal condition (\ref{eq:ns3}) will be satisfied if the normalization of the functions $ \chi ^{\alpha} (r) $ obeys
\begin{align}
 \lambda ^{\alpha} \int \limits_0 ^ 1 dr \, r ^ 2 \, g (r) \, [\chi ^{\alpha} (r)] ^ 2 = -1 \ .
\label{eq:ns9}
\end{align}

In particular, for $ g (r) = - 1 $ we obtain the well-known biorthogonal basis:
\begin{align}
 \chi_l ^{\alpha} (r) = \frac{\sqrt{2}}{\varkappa_{\alpha}} \,
 \frac{1}{| J_{l + 1/2} (\varkappa_{\alpha}) |} \,
 \frac{J_{l + 1/2} (\varkappa_{\alpha} \, r)}{\sqrt{r}} \ ,\\ \rho_l ^{\alpha} (r) = - \frac{\sqrt{2} \, \varkappa_{\alpha}}{| J_{l + 1 / 2} (\varkappa_{\alpha}) |} \, \frac{J_{l + 1/2} (\varkappa_{\alpha} \, r)}{\sqrt{r}} \ ,
\label{eq:oldbf}
\end{align}
 where $ \varkappa_{\alpha} $ are the positive roots of $ J_{l-1/2} (\varkappa_{\alpha}) = 0 $, $ \alpha = 1, 2, ... $ (Polyachenko, Shukhman 1981).

Choice of $ g (r) = \rho_0'(r) / \Phi_0' (r) $ is interesting for calculating the shear zero mode $ l = 1 $, with the perturbed potential and density, respectively, $ \chi (r) = \Phi_0'(r) $, and $ \Pi (r) = \rho_0' (r) $. The expansion for (\ref{eq:bf_expansion}) contains only one term corresponding to the minimum eigenvalue $ \lambda = \lambda ^{\alpha = 1} = 4 \pi $ (i.e., $ C ^ 1 = 1 $; $ C ^ k = 0 $ for $ k> 1 $) and the basis function $ \chi ^{1} (r) = \Phi'_0(r) $.

Note that Bertin et al. (1994) discuss a similar technique for infinite systems. However, there is no attempt to build a system so that the expansion of the potential for shear mode contains a single function (which is useful for tests).

The eigenvalue problem (\ref{eq:ns4}) in the form of a differential equation can be reduced to a more convenient eigenvalue problem in the form of an integral equation. In doing so, we write the equation (\ref{eq:ns4}) in the equivalent form:
\begin{align}
  \chi(r)=-\frac{\lambda}{2l+1}\int\limits_0^1 dr' r'^2\,
g(r')\,\chi(r')\,{\cal F}_l(r,r')\ ,
\label{eq:ns15}
\end{align}
where ${\cal F}(r,r')= {r_<^l}/{r_>^{l+1}}$; $r_<={\rm min}(r,r')$,  $\ \ r_>={\rm max}(r,r')$. In this form, the boundary conditions are satisfied {\it automatically}.

Note that construction of a basis set for radial perturbations ($ l = 0 $)
 is not covered by above analysis, and should be discussed separately.

\label{lastpage}
\end{document}